# Neural Network-Based Feature Extraction for Multi-Class Motor Imagery Classification

Souvik Phadikar, Nidul Sinha, Rajdeep Ghosh

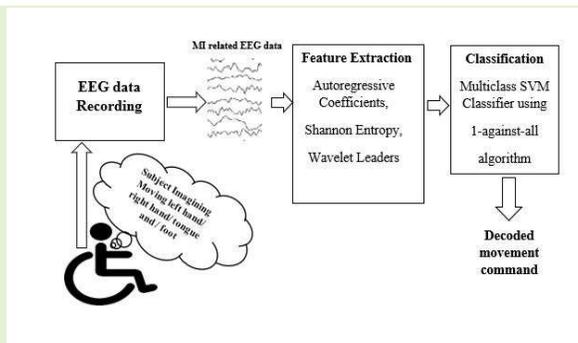

***Abstract*—** Decoding of motor imagery (MI) from Electroencephalogram (EEG) is an important component of the Brain-Computer Interface (BCI) system that helps motor-disabled people interact with the outside world via external devices. The main issue in developing the EEG based BCI is the informative confusion due to the non-stationary characteristics of EEG data. In this work, an innovative idea of transforming an EEG signal into the weight vector of an unsupervised neural network called the autoencoder is proposed for the first time to solve that problem. Separate autoencoders are trained for the individual EEG data. The weight vectors are then optimized for the individual EEG signals. The EEG signals are thus represented in a new domain that is in the form of weight vectors of the individual autoencoder. The weight vectors are then used to extract features such as autoregressive coefficients (ARs), Shannon entropy (SE), and wavelet leader. A window-based feature extraction technique is implemented to capture the local features of the EEG data. Finally, extracted features are classified using a classifier network. The proposed approach is tested on two publicly accessible EEG datasets (BCI competition-III and Competition-IV) to ensure that it is as successful as and superior to the previously published methods. The proposed technique achieves a mean accuracy of 95.33 % for dataset-IIIa from BCI-III and a mean accuracy of 97% for dataset-IIa from BCI-IV for four-class EEG-based MI classification. The experimental outcomes show that the proposed approach is a promising way to increase BCI performance.

## I. Introduction

BCI is a collaborative setup between the brain and an external device, which takes brain signals as input and tries to decode into computer commands to direct external activities such as cursor control, wheel chair control, silent speech recognition, etc. Several techniques are used to capture the neuronal activity inside the brain. EEG is one of them and is widely used because of its non-invasive nature and high resolution in time. It uses electrodes on the scalp to capture the electrochemical changes in the brain. The MI classification through EEG is one of the widely used BCI applications. The cerebral activities of MI can be triggered when a person imagines any movement of his body parts. If these cerebral activities are properly translated, then the findings can be utilized to interact with the external equipments such as BCI-based wheelchairs for physically challenged people and service robots for several motor-neuron diseases (i.e., Poliomyelitis, Parkinson disease, etc.) [1]. The detection of EEG patterns is thus crucial in BCI applications for MI.

In the MI-based BCI systems, unique aspects of the EEG signals called features are extracted and are then integrated into a feature array. Finally, these arrays are further used to train the classifier network. The most important key factor to achieve high performance in BCI systems (higher classification accuracy) is the discriminative feature extraction. Good classification accuracy will be achieved if proper features are extracted for the same. Three major factors such as artifacts, non-stationarity, and misrepresentation of training feature sets can degrade the accuracy of the BCI systems. Electro-oculogram (EOG) and myogenic artifacts are unavoidable contaminations recorded together with neural activities and thereby distort valuable information [2]. Variations in different neurophysiological circumstances of subjects while recording the EEG can cause immense non-stationarity in the EEG signals [3]. The improper imagination of mental tasks and inappropriate class labeling results in deviation in the training data. Various advancements in feature extraction techniques and classification algorithms have been created in the study to address these difficulties. The major goal of this work is to improve feature extraction approaches in order to improve the performance of multi-class MI task classification.

For decoding MI from EEG data, the common spatial pattern (CSP) is commonly employed as a feature extractor [4 - 7]. Initially, it was developed for binary data classification and then

S. Phadikar is with Electrical Engineering Department, National Institute of Technology Silchar, Assam, India (e-mail: souvik_rs@ee.nits.ac.in).

N. Sinha is with Electrical Engineering Department, National Institute of Technology Silchar, Assam, India (email: nidul.sinha@gmail.com).

R. Ghosh is with Information Technology Department, Gauhati University, Assam, India (email: rajdeep.publication@gmail.com).

modified for multi-class classification problems. In conventional CSP analysis, the data of two classes are spatially filtered to maximize the variance between the classes. EEG data are bandpass filtered between a frequency band of 4 to 40 Hz before the CSP is used to decode the MI. The frequency band chosen for CSP has a significant impact on its performance. It is therefore necessary to select a specific frequency band for extracting specific characteristics, but this process is very inconvenient. When the frequency band is used inappropriately, the BCI system's performance suffers. Several CSP extensions have been proposed in the literature to address this issue, including filter bank common spatial pattern (FBCSP) [4], Separable Common Spatio-spectral Patterns (SCSSP) [5], and one-vs-rest FBCSP [6]. However, extended-CSP-based methods are mainly suitable for stationary signals, and considering it for non-stationary EEG signals is quite difficult [8]. Ghosh *et al* [25] proposed a heuristically optimized CSP for classifying the MI tasks from the BCI competition dataset.

Bhattacharya *et al.* [9] proposed adaptive autoregressive (AAR) model-based feature extractor for multi-class MI task classification. However, their methods didn't achieve higher accuracy as AAR based method ignores the temporal features of EEG signals. To capture the time-frequency features of EEG signals, wavelet transform based features extractors have been proposed in the literature [10 - 12]. Rashid *et al.* [10] proposed DWT based feature extraction technique, where, wavelet coefficients at various levels were used as features and provided to the neural network for classifying multi-class MI tasks from EEG signals. Mahamune *et al.* [11] proposed CWT based feature extraction technique. In their approach, 2-D images were formed using wavelet coefficients and used as features in convolutional neural network (CNN) for four-class MI task classifications. Ma *et al.* [12] extracted power spectral density (PSD) from the DWT coefficients of EEG signals and converted them into 2-D images. They used 2-D images in CNN for classifying four-class MI tasks and achieved higher classification accuracy. It is difficult to use wavelet coefficients directly as signal features because of their wider length. However, challenges that remain in wavelet transform-based signal analysis are the selection of appropriate mother wavelet and decomposition level.

Recently, Shi *et al.* [8] combined feature extraction techniques to improve classification accuracy. In their method, CSP and AAR features were extracted and tested on various classifier networks. Zhang *et al.* [13] proposed LASSO-based feature selection methods to progress the performance of decoding MI tasks. The performance was improved through selecting significant features from the extracted features: AR, band power, and wavelet coefficients.

From the previous research, it is observed that the CSP and wavelet transform were widely used as feature extractors in the four-class MI classification. The commonly used CSP algorithm, however, recognizes only spatial-based features while paying no attention to the spectral properties of EEG signals. The performance depends on the spectral filter for which the frequency band is usually predetermined and fixed manually. In AAR-based approaches, time information is ignored. In wavelet transform based techniques, considering coefficient vectors at different level as a feature of EEG signal increase the computational complexity of the system. However, challenges remain in achieving higher classification accuracy to improve the performance of real-time MI-based BCI.

It is quite difficult to decode four-class MI tasks from the EEG data because of its non-stationary nature. Hence, the derived features may not be entirely discriminative. To address the challenge of finding discriminatory features for four-class MI classification because of highly random and non-stationary EEG signal, an innovative idea of transforming the EEG signal in a new domain i.e., weight vector of unsupervised neural network has been proposed in this paper for four-class MI classification for the first time. An efficient feature extraction method is developed rather than improving the classification algorithm to enhance the performance of BCI.

In view of the above challenges, the main features of the innovative method proposed in this paper are:
(a) It is a fully automated, unsupervised, and data-driven feature extraction method using individual autoencoders for every EEG signal and does not require any prior knowledge.
(b) The proposed novel feature extraction technique can adaptively capture the intention of motor movements from the EEG data.
(c) It extracts the discriminative feature sets in a new domain in which the classifier can achieve higher classification accuracy for four-class MI data.

A new hypothesis is proposed, tested, and verified in this paper. The hypothesis is to represent the EEG signal in terms of a weight vector from the input layer to the hidden layer of an autoencoder and subsequently use the weight vector for classifying the EEG signals. Separate autoencoders are used for every EEG signal and weight vectors of the autoencoders are optimized according to the individual EEG fed to a corresponding autoencoder. It is to be noted that separate autoencoders are used for every EEG data. For example, if there are 100 EEG signals, then 100 autoencoders are trained for each EEG signals. Hence, the individual signal is represented as optimized weight vector derived during the process of training the corresponding autoencoder. The weight vectors depict a compact representation of the EEG signal. The feature values can then be extracted from a particular weight vector that represents a particular EEG signal in a new domain.

To test the hypothesis, an autoencoder (a type of neural network) is used to find the unique weight vector for each MI EEG signal. The autoencoder has been used in this paper as it reconstructs the same output as the input. Hence, it is a data-driven network. As a result, the weight vectors are updated according to the individual EEG data. Once the weight vectors are computed, features are then extracted through a slid-windowing method. The windowing-based feature extraction technique is implemented to capture the local features of EEG data and to reduce the dimension of the feature vector. The window size is selected through the proper technique. Then all the windowed feature values are concatenated to form a feature vector. Finally, these feature vectors along with class labels are fed to the SVM network as a classifier. Once the training of the classifier network is completed, individual autoencoders are created for the test signals. Optimized weight vectors representing the individual EEG signals are obtained from the autoencoders. Then, the feature values are extracted and class labels are predicted from the pre-trained SVM network. This method is fully automatic and unsupervised. The proposed system is compared with the conventional MI-based BCI systems [11 - 13] and also achieves the highest classification accuracy. The proposed system is validated with two publicly available datasets. The proposed system neither uses any feature selection techniques nor any channel selection method for achieving higher accuracy than the conventional methods. To the best of our knowledge, this is the first attempt to represent EEG signals as the weight vector for the EEG-based MI classification. The benefits of the stated model are established through the exploratory outcomes as explained in this paper.

The following is a summary of the manuscript: Section I introduces the background, literature review, challenges, and objectives; Section II depicts the tools and ideas used in this methodology; Section III states the proposed methodology; Section IV discusses the research findings; and Section V concludes the manuscript by discussing future work strategies.

## II. MATERIALS AND TOOLS

### A. Autoencoder

An autoencoder is a neural network that captures the signature within EEG data using an unsupervised machine learning technique [14]. It consists mostly of two parts: an encoder and decoder. Encoder converts the input signal into a code and then the code is again converted back into the original input through decoder. Weights between layers are updated according to the training algorithm to minimize the reconstruction error. As the reconstruction error is small enough, the code could be assumed to incorporate most of the information of the input vector. The architecture of an autoencoder is shown in figure 1.

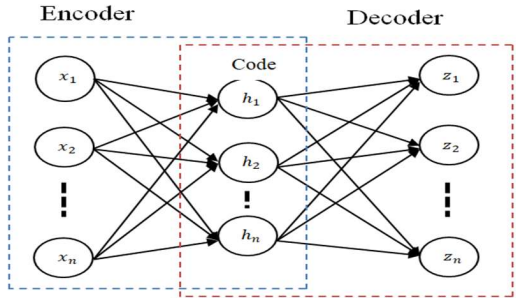

Figure 1: The architecture of autoencoder. $X_i$, $h_i$ and $Z_i$ denote the input node, hidden nodes, and output nodes respectively.

Autoencoder translates the input vector x to a hidden layer representation y using deterministic mapping, as shown in equation (1):

$$y = f_{encoder}(W_e^T x + b_e) \quad (1)$$

Where, $W_e$ and $b_e$ is the weight and bias vector respectively of the encoder and $f_{encoder}$ is the activation function of the neurons in the encoder. The output of the autoencoder z which has the same phase as x is then extracted by mapping the hidden layer representation or code *y* using the transformation. The z is expressed as equation (2):

$$z = f_{decoder}(W_d^T y + b_d) \quad (2)$$

The weight vector and bias vector for the decoder are denoted by $W_d$ and $b_d$, respectively. The activation function of neurons in the decoder is $f_{decoder}$. Training may be done in an autoencoder network by reducing the reconstruction error, which is quantified as a squared error. The autoencoder finds the appropriate parameters $\theta = \{W_e, W_d, b_e, b_d\}$ through minimizing the cost function as shown in equation (3):

$$E(\theta) = L(x, z) + \lambda \parallel W \parallel^2$$
$$= \sum_i^n \parallel x_i - z_i \parallel^2 + \lambda(\parallel W_e \parallel^2 + \parallel W_d \parallel^2) \quad (3)$$

Where $\lambda \parallel W \parallel^2$ is a regularised parameter that minimizes the L2 norm of parameters to avoid over-fitting, and $L(x, z)$ is the reconstruction error. The weight vector $W = W_e(:)$ can represent the MI signal in new space.

### B. Feature Extraction

Despite the fact that different feature extraction approaches for MI EEG classification have been established, finding an appropriate method for greater classification accuracy remains a challenging job for successful EEG classification. Finding a good feature set for a binary classification task may not be so difficult but for a complex multi-task classification, getting a discriminative feature set is a challenging task. In the proposed work, three features combining time-frequency and spectral information are extracted from the weight vector for the corresponding EEG data.

#### 1) Autoregressive Coefficients (AR)

The AR model is a signal processing representation of a sort of random process that is used to characterise time-varying processes [15]. In a parametric method, it estimates the PSD of an EEG. Therefore, there are no chances of spectral leakage. Hence, it yields better frequency resolution. A linear combination of p prior values of the same signal can be used to describe the signal. Equation (4) can be used to describe the signal x[n] at time instant n:

$$x[n] = -\sum_{i=1}^{p} a[i]x[n-i] + e[n] \quad (4)$$

Where e[n] is a white noise, $a[i]$ is the $i^{th}$ coefficient of the model with order p. A total of p number of AR coefficients are used as feature values in this work. However, the selection of the order p is very sensitive because the estimates generally improve with an increase in order but requires a higher computational cost.

- **Burg's Method to Estimate the AR Model**

For the estimate of AR models, several techniques have been presented. In the field of EEG categorization, Burg's technique is frequently utilised. It directly measures the reflection coefficients without using the autocorrelation function [16]. This approach estimates the data records of PSD that exactly resemble the original signal. For a detailed description, interested readers may refer to [17].

- **Determination of AR Model Order**

It's critical to establish the model's order that best suits the data while creating an AR model. It depends on the data sampling rate because the AR model estimates the present value of data using some past data samples. Sum-squared error (SSE) is a widely used tool for determining the order of an AR model. The model with the lowest SSE is the one that best matches the data [18]. As suggested in [18], for EEG-based mental state classification, AR coefficients of order 6 best fit the data.

2) *Wavelet Packet Entropy*

The wavelet transform is widely used in feature extraction due to its capability of capturing the time-frequency features of an EEG signal. It is difficult to use such coefficients directly as features because of their wider length. As a result, certain higher-level features can be derived from these coefficients for improved classification. Entropy is a technique used in information theory and signal processing to quantify the uncertainty of a particular system. Shannon entropy (SE) is computed directly from the weight vector's wavelet packet decomposition (WPD) coefficients in this paper. The main distinction between WPD and DWT is that WPD decomposes both the approximation and detail coefficients at the same time. As a result, the WPD has the same frequency bandwidth in each resolution while DWT does not. For an EEG signal *x(t)*, the coefficients can be derived as equation (5):

$$\begin{cases} d_{0,0}(t) = x(t), \\ d_{i,2j-1}(t) = \sqrt{2} \sum_k h(k) d_{i-1,j}(2t - k), \\ d_{i,2j}(t) = \sqrt{2} \sum_k g(k) d_{i-1,j}(2t - k) \end{cases} \quad (5)$$

Where $h(k)$ and $g(k)$ denote high pass and low pass filters respectively, and $d_{i,j}$ is the WPD coefficients at the $i^{th}$ level and $j^{th}$ node. The energy at $i^{th}$ level and $j^{th}$ node can be derived by wavelet-energy, defined as equation (6):

$$E_{i,j} = \sum_{k=1}^{N} \| d_{i,j,k} \|^2 \quad (6)$$

Where N denotes the total number of coefficients in the corresponding node. The Shannon entropy (SE) of $j^{th}$ node at $i^{th}$ level is calculated based on the probability distribution of energy as equation (7):

$$SE_{i,j} = - \sum_{k=1}^{N} P_{i,j,k} * \log(P_{i,j,k}) \quad (7)$$

Where, $P_{i,j,k}$ is the probability of the $k^{th}$ coefficient at its corresponding node and is defined as equation (8):

$$P_{i,j,k} = \frac{\| d_{i,j,k} \|^2}{E_{i,j}} \quad (8)$$

Finally, the SE feature vector is computed by cascading all the SEs from every node of level M as equation (9).

$$SE = (SE_{i,1}, SE_{i,2}, \dots, SE_{i,2^M})_{i=M} \quad (9)$$

- **Selection of Base Wavelet**

The selection of an appropriate base wavelet (mother wavelet) may affect the calculation of the SE feature vector in the wavelet domain. Hence, a cross-correlation-based approach is proposed to check the performance of all the available wavelet bases for EEG-based MI signal classification. The cross-correlation between the MI signal and the wavelet functions is calculated and the function is selected which gives the maximum value. The correlation $X_{corr}$ between the EEG signal of interest $X$ and the mother wavelet function $Y$ is calculated as equation (10):

$$X_{corr} = \frac{\sum (X - \bar{X})(Y - \bar{Y})}{\sqrt{\sum (X - \bar{X})^2 (Y - \bar{Y})^2}} \quad (10)$$

3) *Wavelet Fractal Estimates*

Two fractal parameters from DWT coefficients are estimated and used as features. The second cumulant of the scaling exponents and the width of the singularity spectrum are derived as features. The width of the singularity spectrum derives from the multi-fractal nature of the EEG signal. The scaling exponents are scale-based exponents that describe the signal's power-law behaviour at various resolutions. The second cumulant roughly indicates the scaling exponents' deviation from linearity [19]. Both the features are calculated from wavelet leaders. The wavelet leaders estimate the multifractal spectrum based on wavelet transform.

Let $\psi$ be a wavelet function having various null moments and fast decay and dilated by a scale $2^j$ and translated to time position $2^j k$. It can be assumed that, each wavelet coefficient $C_{jk}$ corresponding to the wavelet transform of the series *{x(i)}* is localized on the dyadic interval [19], $I_{jk} = \left[\frac{k}{2^j}, \frac{k+1}{2^j}\right]$. Then the dilated intervals can be computed as equation (11):

$$3I_{jk} = \left[\frac{k-1}{2^j}, \frac{k+2}{2^j}\right] \quad (11)$$

The wavelet leaders $d_{jk}$ are computed as equation (12):

$$d_{jk} = \sup\{|C_{lh}| : I_{lh} \subset 3I_{jk}\} \quad (12)$$

The most important key factor about the wavelet leader in the search for the greatest wavelet coefficients in a narrow time neighborhood for a given time and scale [19]. The singularity spectrum (SS) determines how many singularities are there. However, the SS can be easily computed from the structure-function (SF). The SF is computed from the wavelet leader as equation (13):

$$S(q, 2^j) = \frac{1}{n_j} \sum_{k=1}^{n_j} |d_{jk}|^q \quad (13)$$

The SF decays as power laws of the scales if the signal x(i) displays some sort of self-similarity. Scaling exponents (SE) are the exponents of these power laws, and they are calculated using equation (14):

$$SE(q) = \lim_{j \to 0} \inf \left( \frac{\log_2 \left( S(q, 2^j) \right)}{j} \right) \quad (14)$$

Finally, using the Legendre transform (LT), the SS may be derived from the SE as follows:

$$D(h) = \inf_q (1 + qh - SE(h)) \quad (15)$$

The width of the SS is measured as the difference between the maximum and minimum value in the D(h) and the second cumulant of the SE is used as feature values.

All the feature values are extracted from the autoencoder weight vector using the non-overlapping sliding windowing technique.

### C. Multi-class SVM Classifier

The SVM is the most commonly used classifier based on the supervised machine learning method. Using the kernel technique, SVM can adequately categorise non-linear data. To classify test data, the SVM uses training data to construct an optimum hyperplane [20]. The best hyperplane, also known as support vectors, is used to create a decision boundary using neighbouring samples from various datasets. If the datasets are linearly indistinguishable in the original finite-dimensional space, the non-linearity can be reduced by re-mapping the data onto a suitably higher dimensional space. To address the higher dimensionality, the kernel trick is used for better classification and less effort. However, SVMs were originally designed for binary classification problems. Various extensions of SVM have been proposed in the literature for multiclass classification problems. These extensions are 1-against-1, 1-against-all, DAGSVM etc. [20].

### D. Dataset

In this investigation, two datasets were used: one for validation and another for comparison with other recently reported methods of EEG-based MI Classification.

#### 1) Dataset IIIa from BCI-III

Dataset IIIa from BCI competition III [21] has been used to validate the proposed system. The data consist of records from the three subjects (k3b, k6b, l1b). The subjects performed four MI tasks according to a cue. The subjects were imagining the movements of the left hand (class-1), right hand (class-2), tongue (class-3), and foot (class-4) while relaxing in a chair with armrests. The experiments were performed for a few runs with 40-trials for each class. At the begining, the 2s were silent and an audio stimulus started at t = 2s to indicate the starting of each trial, and across "+" was displayed. Then, at t = 3s, a left, right, up, or down arrow was presented for 1s and at the same time, and the participants were instructed to imagine moving their left hand, right hand, tongue, or foot in accordance with the arrow until the cross vanished at t = 7s. Each of the four cues is shown 10 times in a random order during each run.

#### 2) Dataset IIa from BCI-IV

Dataset IIa from BCI competition IV [22] has been adapted to illustrate the superiority of the proposed model over other recently established techniques. EEG data were collected from 9 individuals (A01–A09) using a 22-channel EEG amplifier. The subjects were imagining the movements of the left hand (class 1), right hand (class 2), tongue (class 3), and foot (class 4) while relaxing in a chair with armrests. The data were sampled at 250 Hz.

### III. PROPOSED METHODOLOGY

This paper proposes a novel method of unsupervised feature extraction for EEG-based MI classification. The basic architecture of the proposed system is presented in figure 2. At first, the MI-related EEG signal is fed to an autoencoder and the weight vector which minimizes the reconstruction error in the autoencoder is extracted. The individual weight vectors are extracted after training the individual autoencoders for each EEG signals. The autoencoders are trained with the EEG data from individual channels in a trial. In the proposed system, the size of the hidden layer is selected as 50 after experimentation and is elaborated in the next section. Thus, the size of the input EEG signal is 750 samples for an individual channel, and hidden node size is 50. Then, the size of weight matrix is 750×50. The weight matrix is then transformed into a vector of size (1×(750×50=37500)). From, the weight vectors, features are extracted as mentioned above. Here, p number of AR coefficients, $2^M$ number of Shannon entropy and 2 number of wavelet fractal estimates are extracted from each window. Finally, (p + $2^M$ + 2) * S number of feature values are concatenated to form the feature vector, where S is the total number of windows to cover the whole weight vector. Finally, the feature-label pairs are fed to a multiclass SVM classifier to train the model. The proposed method is implemented in MATLAB R2019a and evaluated for BCI-III and BCI-IV datasets. The proposed work achieves a higher classification accuracy as compared to the conventional methods ([11 - 13]) reported for EEG-based MI classification.

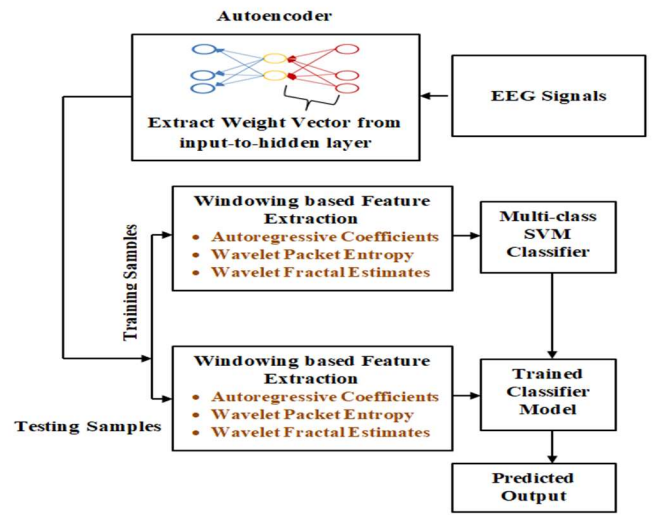

Figure 2: The basic architecture of the proposed system.

**Stages of the proposed model**
➢ Four class MI data were segmented from the dataset.

- Data were partitioned into a training set and test set using the CV partition technique with holdout parameter p = 0.3. 70% of the data were selected randomly for the training purpose and 30% of the data were selected randomly for the testing purpose.
- Individual EEG data from the training set are fed to train the individual autoencoders. After minimizing the training error of the individual networks, the weight vector from the input-to-hidden layer is extracted. This weight vector represents the corresponding EEG data from each class.
- The features are extracted from the weight vector according to a sliding window methodology and a window size of 250 yields the best results. From, the weight vectors, features are extracted as mentioned above. Here, p number of AR coefficients, $2^M$ number of Shannon entropy and 2 number of wavelet fractal estimates are extracted from each window. Finally, $(p + 2^M + 2) * S$ number of feature values are concatenated to form the feature vector. Likewise, test feature sets were also extracted using the same method.
- Finally, the training feature-label pairs were fed to a multiclass SVM to train the classifier network.
- To test the trained classifier network, test feature sets are used and the performance is evaluated.

### A. Data pre-processing and preparation

The EEG data is very sensitive to various artifacts such as eyeblink, eye movement, etc. These artifacts should be removed otherwise result in misclassification. In this paper, the EEG artifacts are removed through the method described in [23]. At first, the EEG data are decomposed into wavelet coefficients up to level 6 using db8 as the mother wavelet. Then, according to the wavelet coefficients, the appropriate thresholds are calculated through a heuristic algorithm (grey wolf optimization) and coefficients are thresholded. Finally, thresholded coefficients are used in the inverse operation to get back the artifact-free EEG signal.

After successfully removing the artifacts from the BCI-III (data IIIa) dataset, the MI-related EEG data were segmented. EEG segments from 4s to 7s in an individual trial were extracted as it represents the MI task. Therefore, 32 channel EEG data representing an individual MI task within an individual trial were extracted from the available dataset.. Then, data were randomly partitioned into a train ($X_{train}$) and test ($X_{test}$) dataset using the holdout CV algorithm. In the CV partition algorithm, the holdout parameter, p is selected as 0.3 and as a result, 70% of the total dataset were randomly chosen to form a training dataset, and 30% of EEG segments were kept as test datasets. Finally, the data were paired with data labels and are ready to be applied in the proposed method.

### B. Transforming the EEG signals into Weight Vectors

Now, the $X_{train}$ and $X_{test}$ were fed to train the individual autoencoders, and the encoder weight vector for each data segment is stored for further investigation. The activation function of the encoder network is selected as 'logsig' function.

For the given input, $X_i$, the encoder output $y_i$ is calculated as equation (16):

$$y_i^{Train} = logsig(W_i^{Tr} X_{train}(i) + b_e) \quad (16)$$

And, similarly for the test dataset $X_{test}$:

$$y_i^{Test} = logsig(W_i^{Te} X_{test}(i) + b_e) \quad (17)$$

The 'logsig' function is expressed as equation (18):

$$logsig(z) = \frac{1}{1 + e^{-z}} \quad (18)$$

To minimize the error defined in equation (3), the scaled conjugate gradient (SCG) algorithm [24] is used as a training algorithm in autoencoder. The weight vectors are selected, when the autoencoder reconstruction error is minimized. The number of hidden nodes in the hidden layer was selected using the trial and error method. However, EEG data is not linearly separable and it is difficult to estimate how many hidden nodes will be required to represent the EEG data efficiently. Hence the size of hidden nodes is selected as 30, 50, and 100, and the training errors of the autoencoders are compared. It has been observed that the error is the minimum with fast convergence when the number of hidden nodes is 50. Figure 3 represents the training error for an individual EEG on using different hidden node sizes.

Now, the weight vector ($W_i$) for the corresponding EEG signal are extracted.

### C. Feature Extraction from the Weight Vectors

A windowing feature extraction method is proposed in this work. The window slides across the weight vector and the features are computed. Each window will calculate the p number of AR coefficients, $2^M$ number of wavelet packet entropy, and two numbers of wavelet fractal estimates as a feature value from the EEG segments. Hence, each EEG signal will be represented as $(S \times r)$ number of feature values. Where r is the total number of feature values extracted from a window and S is the total number of windows to cover the whole weight vector.

$$r = (p + 2^M + 2) \quad (19)$$

- In computing AR coefficients, the selection of order p is a key factor. As the researchers in [19] suggested that, order 6 is significant in the AR model for mental task classification. Hence, the order of AR model p = 6 is selected in the proposed methodology. Also, Burg's method (described in section II) is used to estimate the AR coefficients due to its low computational cost.
- Wavelet packet entropy, is calculated from an individual window of the weight vector. One major issue in the computation of wavelet-based SE is the selection of proper wavelet function. Several wavelet functions are available in the wavelet families, and every wavelet function has different characteristics. As a result, entropy will be different for different wavelet functions. However, in this work, a proper technique for the selection of appropriate wavelet functions is adopted. The correlation $X_{corr}$ between the EEG signal of interest $X$ and the mother, wavelet function $Y$ is calculated using equation (10) to match the best suitable wavelet function for EEG-based MI analysis. The wavelet function is selected which gives the maximum value. In this work, a total of 30 wavelet functions (Haar, db2 – db12, Coif1 – Coif5, Sym2 – Sym8, fk4 – fk22) are compared with MI-related EEG data. For each wavelet function, the average $X_{corr}$ is shown in table 2. From the table, it is evident that the db4 provides the maximum correlation among all the wavelet functions. Hence, db4 is selected as a mother wavelet function to decompose the windowed EEG data using WPD up to level 4. Hence M = 4, and from the equation (9), $2^M = 16$ wavelet packet entropy will be computed from a window. Finally, the second cumulant of the SE and the width of the SS is measured from the equation (14) and (15) respectively from a window.

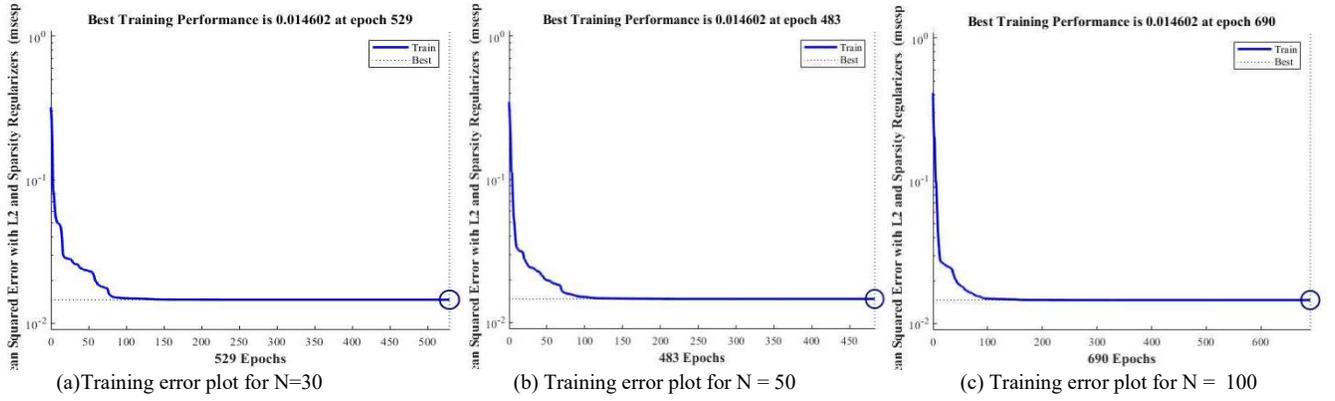

(a) Training error plot for N=30  (b) Training error plot for N = 50  (c) Training error plot for N = 100

Figure 3: Performance of the autoencoder with different configuration

- In total (6 AR coefficients + 16 wavelet packet entropy + 1 SE + 1 SS) = 24 features are extracted from a window. Thus, the size of weight vector is 37,500 and the size of the sliding window is 250. Hence, 37500/250=150 number of windows are required to traverse the entire weight vector. Finally, all the feature values computed from every window are concatenated to form the feature vector, which consists of (150 * 24) = 3600 feature values.

Table 2: Wavelets vs. Average cross correlation.

| Wavelets | Xcorr | Wavelets | Xcorr | Wavelets | Xcorr |
|---|---|---|---|---|---|
| Haar | 0.12 | db11 | 0.02 | coif2 | 0.02 |
| db2 | 0.05 | db12 | 0.02 | coif3 | 0.02 |
| db3 | 0.04 | sym2 | 0.05 | coif4 | 0.02 |
| db4 | 0.04 | sym3 | 0.04 | coif5 | 0.02 |
| db5 | 0.03 | sym4 | 0.03 | fk4 | 0.06 |
| db6 | 0.04 | sym5 | 0.03 | fk6 | 0.04 |
| db7 | 0.02 | sym6 | 0.02 | fk8 | 0.03 |
| db8 | 0.02 | sym7 | 0.02 | fk14 | 0.03 |
| db9 | 0.01 | sym8 | 0.02 | fk18 | 0.01 |
| db10 | 0.02 | coif1 | 0.04 | fk22 | 0.02 |

D. Classification

On the classification stage, a one-against-all SVM classifier is used in this work for four-class MI data classification. Then the feature-label pairs were fed to the classifier to train the network. Additionally, the Gaussian kernel function is used in the classifier for better classification.

IV. EXPERIMENTAL RESULTS AND DISCUSSION

The dataset IIIa from BCI-III has been used to evaluate the proposed methodology in this paper. One trial from each class was selected randomly. After successfully partitioning the data into the training set and test set, both the datasets were fed to the autoencoder. The weight vector of corresponding EEG data was extracted. Then features were extracted through a sliding window from the weight vector. The selection of the size of the window is very sensitive because lower size results in a large number of feature values which may be redundant and a higher size result in a smaller number of feature values which may result in poor classification. The window slides forward across the weight vector. So, depending on the size of weight vectors, the window size is selected as 125, 250, 375, 500, 750, 1000, 1250, and 1500 samples. The performance of the window sizes is compared in figure 4. It is evident from the figure that, the window size of 250 samples has achieved the highest classification accuracy. Finally, the feature-label pairs were fed to the classifier to train the network. After successfully training the classifier, the trained model has been used to predict the test dataset.

Table 3: Comparison among several case studies (on dataset IIIa BCI-III)

| Methods | Classification Accuracy (%) |
|---|---|
| **Case-1** | 85 |
| **Case-2** | 91.87 |
| **Case-3** | 95.39 |

Table 4: Performance comparison of the proposed system on dataset IIIa from BCI-III

| Parameters (%) | LDA | | | | | SVM | | | | |
|---|---|---|---|---|---|---|---|---|---|---|
| | Class-1 | Class-2 | Class-3 | Class-4 | Mean | Class-1 | Class-2 | Class-3 | Class-4 | Mean |
| **Sub_1** | | | | | | | | | | |
| Sensitivity | 100 | 83 | 62 | 69 | 78.5 | 94 | 93 | 94 | 100 | 95.25 |
| Specificity | 89 | 90 | 95 | 97 | 92.75 | 99 | 98 | 100 | 97 | 98.5 |
| Precision | 76 | 73 | 84 | 86 | 79.75 | 97 | 93 | 100 | 90 | 95 |
| Model accuracy | 78 % | | | | | 95 % | | | | |
| **Sub_2** | | | | | | | | | | |
| Sensitivity | 84 | 66 | 97 | 69 | 79 | 100 | 97 | 100 | 96 | 98.25 |
| Specificity | 94 | 96 | 85 | 98 | 93.25 | 99 | 99 | 100 | 100 | 99.5 |
| Precision | 84 | 83 | 72 | 90 | 82.25 | 97 | 97 | 100 | 100 | 98.5 |
| Model accuracy | 80 % | | | | | 98 % | | | | |
| **Sub_3** | | | | | | | | | | |
| Sensitivity | 68 | 93 | 79 | 77 | 79.25 | 97 | 90 | 94 | 92 | 93.25 |
| Specificity | 97 | 82 | 98 | 96 | 93.25 | 98 | 97 | 100 | 97 | 98 |
| Precision | 88 | 63 | 93 | 83 | 81.75 | 94 | 90 | 100 | 89 | 93.25 |
| Model accuracy | 79 % | | | | | 93 % | | | | |

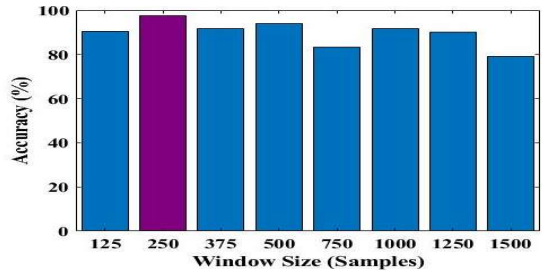

Figure 4: Comparison of different window sizes for feature extraction

Three cases have been studied to check the superiority of the proposed model. In the first case, the same feature sets were extracted directly from the EEG signal and fed to the classifier (case-1). In the second case, features were extracted from the EEG signal (case-2) window-wise. In the third case, features were extracted using the proposed methodology i.e. from weight vectors (case-3) of the autoencoders. For all three cases, the performance is presented in table 3. It is evident from the table that, the classification accuracy using the proposed model (case-3) is higher than the other two cases. The multiclass SVM classifier has been adopted through comparison with other classifiers. Linear Discriminant Analysis (LDA) network also has been used to evaluate the proposed method. Their performance is shown in table 4. From the table, it is evident that the SVM shows the highest performance in terms of precision, sensitivity, specificity, and model accuracy.

The proposed system is also evaluated on the BCI-IV-2a dataset. Table 5 represents the performance of the proposed system on dataset IIa from BCI competition IV. Among all the 9 subjects, subjects: A02, A03, A04, and A08 the prediction is 100% accurate for four-class MI EEG data. The performance of the proposed system is also compared with the other recently reported methods ([11 - 13]) and shown in table 6. It is evident from the table that, the proposed approach gives better results than other methods in terms of classification accuracy.

Table 5: Subject wise performance comparison on dataset IIa from BCI-IV

| Subjects | Classification Accuracy (%) | Average Accuracy (%) |
|---|---|---|
| A01 | 93 | |
| A02 | 100 | |
| A03 | 100 | |
| A04 | 100 | |
| A05 | 97 | 97 |
| A06 | 93 | |
| A07 | 93 | |
| A08 | 100 | |
| A09 | 97 | |

Table 6: Performance comparison of several methods on dataset IIa from BCI-IV

| Performance | Mahamune et al. [11] | Zhang et al. [13] | Ma et al. [12] | Proposed method |
|---|---|---|---|---|
| Average accuracy (%) | 71.25 | 84.96 | 96.26 | 97 |

## V. CONCLUSIONS

In this paper, a new method of feature extraction is proposed for four-class MI EEG classification. The main challenge in the classification of EEG signals is their non-stationary characteristics. To solve this problem, the EEG signal was transformed into another domain i.e., the weight vector of the autoencoder. In the proposed method, the values of three features, AR coefficients, wavelet packet entropy, wavelet fractal estimates were obtained from a new representation of the EEG signals in the form of weight vectors of the autoencoders. The weight vectors used to represent the EEG signals were obtained at the minimum reconstruction error of the autoencoder. A windowing-based feature extraction technique was implemented to capture the local features of EEG signals. Results reveal that an EEG signal can be represented in the weight vector of the autoencoder neural network. The performance of the proposed method was compared with the existing conventional methods. The proposed method successfully predicts the mental task with higher accuracy. It can be concluded that the proposed method can be regarded as a powerful tool to improve the performance of MI EEG-based BCIs.